# A Magnetic Transition Probed by the Ce Ion in Square-Lattice Antiferromagnet CeMnAsO

Yuto TSUKAMOTO[1], Yoshihiko OKAMOTO[1,2], Kazuyuki MATSUHIRA[3], Myung-Hwan WHANGBO[4], and Zenji HIROI[1,2*]

[1]*Institute for Solid State Physics, University of Tokyo, Kashiwa, Chiba 277-8581, Japan*
[2]*TRIP, JST, Sanbancho, Chiyoda-ku, Tokyo 102-0075, Japan*
[3] *Graduate School of Engineering, Kyushu Institute of Technology, Kitakyushu 804-8550, Japan*
[4]*Department of Chemistry, North Carolina State University, Raleigh, NC 27695-8204, USA*

We examined the magnetic properties of the square-lattice antiferromagnets CeMnAsO and LaMnAsO and their solid solutions $La_{1-x}Ce_xMnAsO$ by resistivity, magnetic susceptibility, and heat capacity measurements below room temperature. A first-order phase transition is observed at 34.1 K, below which the ground-state doublet of the Ce ion splits by 3.53 meV. It is likely that Mn moments already ordered above room temperature are reoriented at the transition, as reported for related compounds, such as NdMnAsO and PrMnSbO. This transition generates a large internal magnetic field at the Ce site in spite of the fact that simple Heisenberg interactions should be cancelled out at the Ce site owing to geometrical frustration. The transition takes place at nearly the same temperature with the substitution of La for Ce up to 90%. The Ce moment does not undergo long-range order by itself, but is parasitically induced at the transition, serving as a good probe for detecting the magnetism of Mn spins in a square lattice.

[*]E-mail address: hiroi@issp.u-tokyo.ac.jp

After the discovery of iron arsenide oxide superconductors,[1] a search for related compounds has been extensively carried out, fruitfully resulting in a large family of compounds including many iron-based superconductors.[2] A common playground for superconductivity is a square lattice made up of $Fe^{2+}$ ions, which is to be compared with the square lattice of $Cu^{2+}$ ions in cupric oxide superconductors. In spite of the similarity in their lattice symmetries, the electronic properties of the parent compounds of these superconductors are very different: the former is a semimetal, while the latter is a Mott insulator. In view of the prominent role of the square lattices of $Cu^{2+}$ and $Fe^{2+}$ ions in superconductivity, it is of interest and importance to examine square lattice systems made up of transition-metal ions other than $Cu^{2+}$ and $Fe^{2+}$.

In the present work, we focus on two Mn compounds, CeMnAsO and LaMnAsO, which crystallize in the ZrCuSiAs structure possessing a square lattice made up of $Mn^{2+}$ ions (Fig. 1).

Although the presence of the compounds in nature has been reported in a previous study,[3] their physical properties have rarely been studied.[4] One should expect enhanced magnetism for CeMnAsO and LaMnAsO in comparison with the Fe analogues, because a $Mn^{2+}$ ($d^5$) ion has a higher magnetic moment than the $Fe^{2+}$ ($d^6$) ion. In fact, a recent study on a related compound, $BaMn_2As_2$, crystallizing in the $ThCr_2Si_2$ structure has shown that Mn moments order antiferromagnetically below $T_N = 625$ K with an ordered moment of 3.9 $\mu_B$/Mn at 10 K.[5,6] Moreover, $BaMn_2As_2$ is a semiconductor with a small activation energy of 0.03 eV.[5] It would be interesting to compare the properties of CeMnAsO and LaMnAsO with those of $BaMn_2As_2$ from the perspective of the dimensionality of the magnetic interactions, because ZrCuSiAs-type compounds are more two-dimensional in electronic structure than $ThCr_2Si_2$-type compounds.

Very recently, the magnetic properties of two isomorphic compounds, NdMnAsO and PrMnSbO, have been reported.[7-10] Both compounds undergo an antiferromagnetic long-range order (LRO) of Mn moments directed along the $c$-axis at high temperatures of $T_N = 359$ and 230 K, respectively. In addition, spin-reorientation (SR) transitions are observed at $T_{SR} = 23$ and 35 K, respectively, where the Mn moments lie on the $ab$ plane. Since the Nd and Pr moments appear simultaneously below $T_{SR}$, it has been suggested that their LRO is coupled with the SR transition of the Mn moments. However, the temperature evolution of the Nd or Pr moments determined by neutron diffraction experiments remains unexplained and far from that expected for the conventional magnetic order. Moreover, in NdMnAsO, another anomaly is present in magnetic susceptibility at 4 K, but not in heat capacity.[7] We believe that the present study of CeMnAsO and LaMnAsO with less or nonmagnetic ions instead of Nd and Pr will provide us with important information about the origin of the SR transition and also about basic interactions between $3d$ and $4f$ spins in this class of compounds.

Another motivation for our investigation of CeMnAsO is to study the magnetism of the $Ce^{3+}$ ion. In CeFeAsO, Ce moments undergo an antiferromagnetic order below $T_N(Ce) = 3.7$ or 4.4 K through RKKY interactions.[11-13] In contrast, CeFePO is a magnetically disordered heavy-fermion metal with a large Sommerfeld coefficient $\gamma = 700$ mJ $K^{-2}$ $mol^{-1}$.[14] This difference comes from the fact that the Ce $4f$ state is located well below the Fermi level in the case of CeFeAsO, but just below the Fermi level in CeFePO.[15] Thus, the mixing of the Ce $4f$ state with the states around the Fermi level is negligible in CeFeAsO but substantial in CeFePO. In addition, it is known that the antiferromagnetic order and spin dynamics of the Fe moments in CeFeAsO can be detected by measuring the crystalline electric field (CEF) excitations of the $Ce^{3+}$ ion from inelastic neutron scattering experiments.[16] The $Ce^{3+}$ CEF levels in CeFeAsO are composed of three Kramers doublets at 0, 18.63, and 67.67 meV in the paramagnetic state. When the antiferromagnetic order of Fe moments sets in at $T_N(Fe) = 140$ K, each doublet splits into two singlets. The splitting of the lowest doublet is 0.93 meV. Then, Ce moments exhibit a long-range order below $T_N(Ce) = 4$ K. Therefore, the Ce ion serves as a probe for the magnetism of the Fe ion in the square lattice, and it is of interest to carry out a similar investigation of CeMnAsO.

A polycrystalline sample of CeMnAsO was synthesized by a solid-state reaction of CeAs and $Mn_2O_3$ in an evacuated silica tube at 1000 °C for 48 h in the presence of excess Ti powder used as an oxygen getter. CeAs was prepared by reacting Ce and As powders at 500°C for 24 h and at 850 °C for 5 h. LaMnAsO as well as solid solutions between CeMnAsO and LaMnAsO were also prepared by the same method, starting from a mixture of CeAs and LaAs. Examinations by powder X-ray diffraction analysis with Cu-K$\alpha$ radiation showed that all the samples contained very little impurity phases. Moreover, the lattice parameters of the solid solutions varied systematically, showing complete miscibility between CeMnAsO and LaMnAsO. Magnetic susceptibility, heat capacity, and resistivity measurements were carried out with a Magnetic Property Measurement System and a Physical Property Measurement System (Quantum Design).

Figure 1 shows the electrical resistivities of CeMnAsO, $La_{0.8}Ce_{0.2}MnAsO$, and LaMnAsO as functions of temperature. The former two compounds are insulators with activation energies of 0.13 and 0.14 eV, respectively, estimated from the Arrhenius plots. These activation energies are much higher than that of $BaMn_2As_2$ (0.03 eV).[5] This is consistent with the reported semiconducting character of NdMnAsO[7] but is in contrast to the metallic behavior observed for PrMnSbO.[8] On the other hand, the resistivity of LaMnAsO shows a weak temperature dependence, which can be ascribed to a lower activation energy or to doping effects arising from certain nonstoichiometry in the chemical composition. As will be presented below, these compounds resemble each other in magnetic properties, irrespective of the wide variety of transport properties.

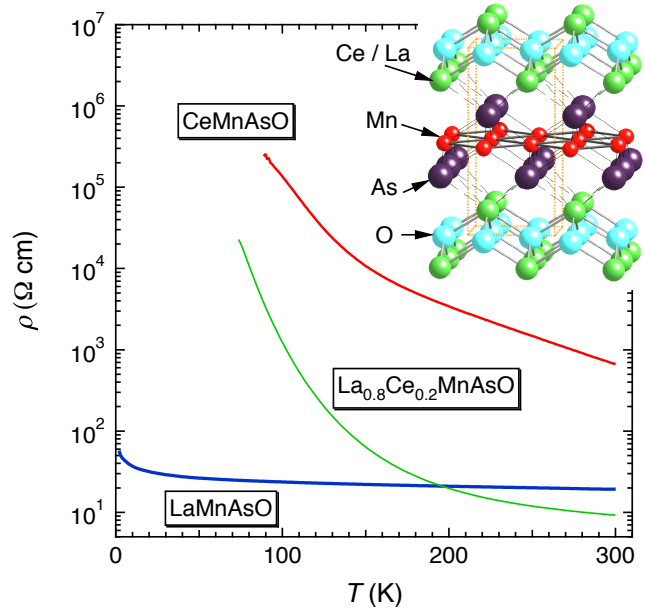

Fig. 1. (Color online) Resistivities of polycrystalline samples of CeMnAsO, $La_{0.8}Ce_{0.2}MnAsO$, and LaMnAsO. The inset depicts the crystal structure of CeMnAsO in the ZrCuSiAs structure.

The magnetic susceptibilities of LaMnAsO and CeMnAsO exhibit different behaviors, as shown in Fig. 2. The susceptibility of LaMnAsO has a weak temperature dependence in a wide temperature range below 300 K, and a large value of $1.23 \times 10^{-3}$ cm$^3$ mol$^{-1}$ at 300 K. This indicates that Mn moments are already ordered above 300 K, which has been confirmed in a recent neutron diffraction study ($T_N$ = 317 K).[10] In contrast, the magnetic susceptibility of CeMnAsO shows a strong temperature dependence that approximately follows the Curie-Weiss (CW) law at high temperatures. By fitting the susceptibility data between 100 and 300 K to the form $\chi = \chi_0 + C/(T - \Theta)$, where $\chi_0$ is a temperature-independent susceptibility, $C$ the Curie constant, and $\Theta$ the Curie-Weiss temperature, we obtain $\chi_0$ = 1.40(2) × $10^{-3}$ cm$^3$ mol$^{-1}$, $C$ = 0.807(9) cm$^3$ K mol$^{-1}$, and $\Theta$ = -28 (1) K. The magnitude of the effective moment deduced from $C$ is 2.54 $\mu_B$, which is exactly the value expected for a free $Ce^{3+}$ ion. Thus, the contribution of the Mn moments must be almost temperature-independent in this temperature range and must be included in the $\chi_0$ term, suggesting that they are already ordered at a higher temperature, as in the cases of LaMnAsO, NdMnAsO, and PrMnAsO with $T_N$(Mn) values much higher than 300 K.[7,8]

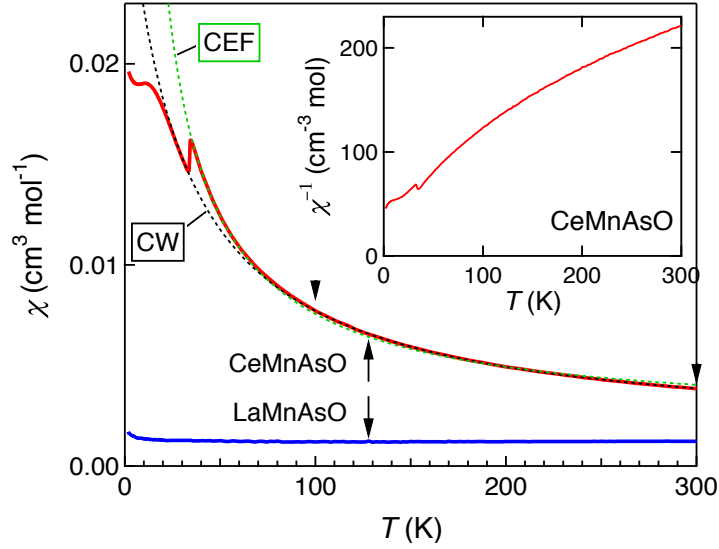

Fig. 2. (Color online) Magnetic susceptibility $\chi$ measured at a magnetic field of 1 T upon heating polycrystalline samples of CeMnAsO and LaMnAsO. The two broken curves represent fittings to the Curie-Weiss (CW) law and were obtained by assuming crystalline electric field (CEF) splittings into three doublets for a $Ce^{3+}$ ion. The data between 100 and 300 K are used for these fittings. Inverse magnetic susceptibility is plotted in the inset.

There is, however, a deviation discernible from the CW law below ~80 K for CeMnAsO. A more elaborate fitting has been performed assuming a CEF level splitting, as in the case of CeFeAsO.[16] Judging from the CW behavior at high temperatures, it is clear that the ground state of the $Ce^{3+}$ ion is a Kramers doublet with no or negligible splitting and that the second doublet is located far above. One would expect CEF splittings similar to those of CeFeAsO because of the isomorphic structure. We have fitted the data between 100 and 300 K by assuming the first excitation energy $\Delta_1$ as a variable and the second one $\Delta_2$ = 67.67 meV as observed in CeFeAsO,[16] and obtained a better fit above 34 K than the CW fit, which yields $\Delta_1$ = 15.9(1) meV and $\chi_0$ = 2.21(7) × $10^{-3}$ cm$^3$ mol$^{-1}$. The $\Delta_1$ for CeMnAsO is slightly smaller than that for CeFeAsO (18.63 meV). Note that the ground-state doublet remains intact even though the Mn moments have already become ordered above room temperature, which is different from the case of CeFeAsO, for which the CEF level splitting takes place once the Fe moments order below 140 K.

Next, we will focus on the low-temperature behavior in $\chi$. There is a cusp in $\chi$ at 34 K, where the magnitude suddenly drops by approximately 10%. Then $\chi$ shows a broad maximum at approximately 13 K, followed by another small increase toward $T = 0$; the last one must be the contribution of a small amount of impurity spins. In contrast, there is no such feature in LaMnAsO. Since $\chi$ is definitely dominated by Ce moments in CeMnAsO, these observations suggest the occurrence of a significant change in the electronic state of Ce at 34 K.

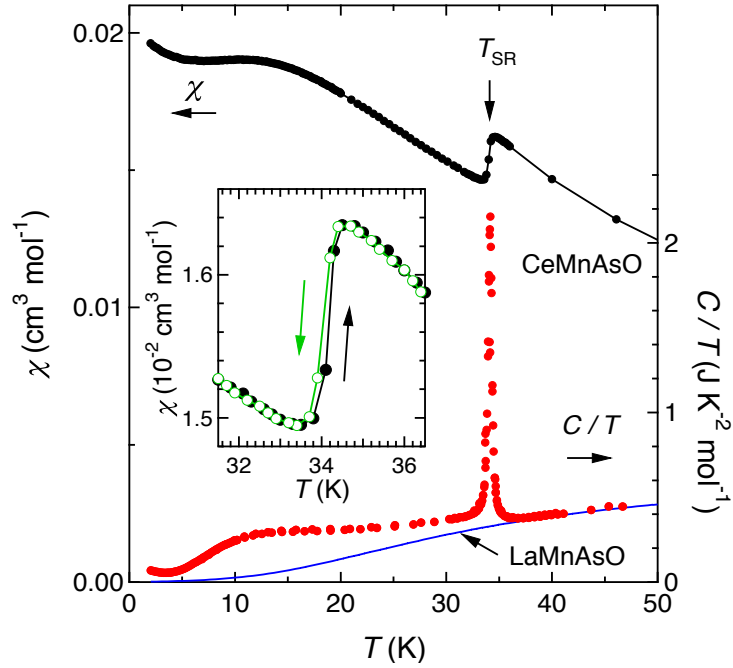

Fig. 3. (Color online) Phase transition at 34.1 K revealed by magnetic susceptibility measured at $\mu_0 H$ = 1 T and heat capacity measured at zero magnetic field. The inset is an enlargement of the cusp observed in $\chi$ measured upon heating and cooling.

The magnetic susceptibility and heat capacity data near the cusp are presented in Fig. 3. As shown in the inset, the magnetic susceptibility shows a small thermal hysteresis between the cooling and heating curves at approximately 34.1 K, suggesting the presence of a weak first-order transition. Correspondingly, the heat capacity exhibits a sharp peak with a peak top at $T_{SR}$ = 34.1 K, which is equal to the temperature at the midpoint of the drop in $\chi$. We cannot detect latent heat, which is characteristic of a first-order transition, in our experimental setup with the relaxation method. However, the peak shape is much different from that expected for a second-order transition. These findings are evidence of a weak first-order phase transition at $T_{SR}$ in the bulk nature. Moreover, there is no change in the basal line of the heat capacity across $T_{SR}$, which suggests that there is no accompanying structural change (or it might be subtle, if any). Below $T_{SR}$, a broad peak appears at approximately 13 K, which obviously corresponds to that observed in the magnetic susceptibility data. In contrast, LaMnAsO shows no corresponding anomaly.

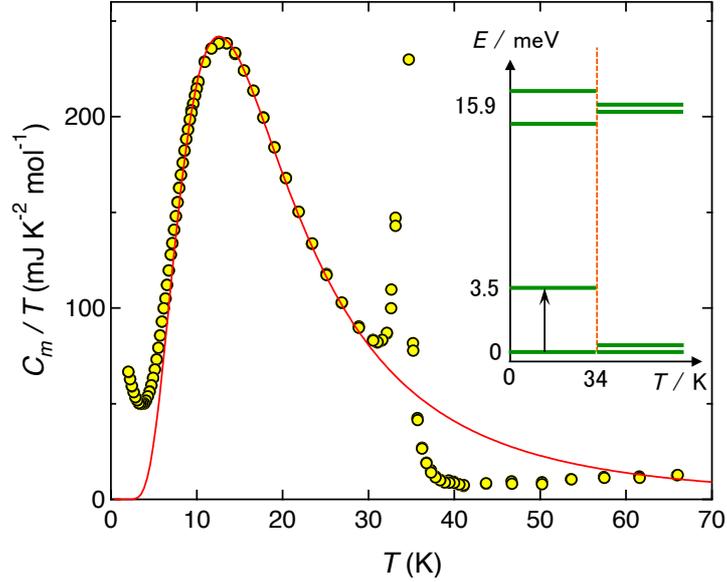

Fig. 4. (Color online) Magnetic heat capacity divided by temperature, $C_m/T$, measured at zero magnetic field for the Ce ion, obtained by subtracting the heat capacity of LaMnAsO from that of CeMnAsO. The solid curve represents a fit to the Schottky-type heat capacity with an energy gap of 3.53 meV. The inset schematically shows $Ce^{3+}$ CEF levels above and below the transition temperature of $T_{SR}$ = 34.1 K: the ground-state doublet splits by 3.53 meV below $T_{SR}$, and the first-excited doublet is located at 15.9 meV above $T_{SR}$, as evidenced by the fitted magnetic susceptibility data shown in Fig. 2.

The magnetic heat capacity of CeMnAsO arising only from the Ce moments, estimated from the heat capacity of CeMnAsO by subtracting that of LaMnAsO, is plotted in Fig. 4. The broad peak bears resemblance to a simple Schottky-type heat capacity if one ignores the upturn toward $T = 0$, which is due most likely to impurity spins, as also observed in magnetic susceptibility. Thus, it is reasonable to attribute the broad peak to the level splitting of the ground-state doublet of the $Ce^{3+}$ ion. Fitting the heat capacity data between 8.4 and 25 K leads to $\Delta$ = 3.53(1) meV ~ 41 K with a scale factor of 0.965(3) (3.5% impurity spins may exist in the sample). The fit is almost perfect below 30 K, but cannot reproduce the data above $T_{SR}$: the heat capacity above $T_{SR}$ is much smaller than the fitted curve, clearly indicating that the splitting is absent above $T_{SR}$ and appears suddenly at $T_{SR}$. The entropy change associated with the sharp peak at $T_{SR}$ is approximately 1.1 J $K^{-1}$ $mol^{-1}$, and that below the peak is 4.7 J $K^{-1}$ $mol^{-1}$, with the sum of the two contributions nearly equal to $R\ln2$ = 5.76 J $K^{-1}$ $mol^{-1}$. Therefore, the total entropy associated with the Ce doublet is partly lifted at $T_{SR}$ and vanishes gradually upon cooling without further transitions.

Figure 5 shows the magnetic susceptibility and heat capacity of a series of solid solutions between CeMnAsO and LaMnAsO. On increasing the La content, the anomalies at $T_{SR}$ become smaller but are always observed at nearly the same temperature with the substitution of La for Ce up to 90%. In contrast, they are absent for pure La compound, suggesting that the transition tends to disappear as $x$ approaches zero. The heat capacity of $La_{0.8}Ce_{0.2}MnAsO$ shows a tiny but sharp peak at 34 K and a reduced broad peak at low temperatures. The magnetic entropy obtained by subtracting the data of the La compound is 1.06 J K$^{-1}$ mol$^{-1}$, which corresponds to 18% of $R\ln 2$. Thus, almost all the magnetic entropy from the 20% Ce ions has been detected. Since the Ce sublattice is highly diluted by nonmagnetic La ions, it is clear that there is no LRO of Ce moments at $T_{SR}$. The Ce ion serves as a good probe for the change in the magnetism of Mn layers.

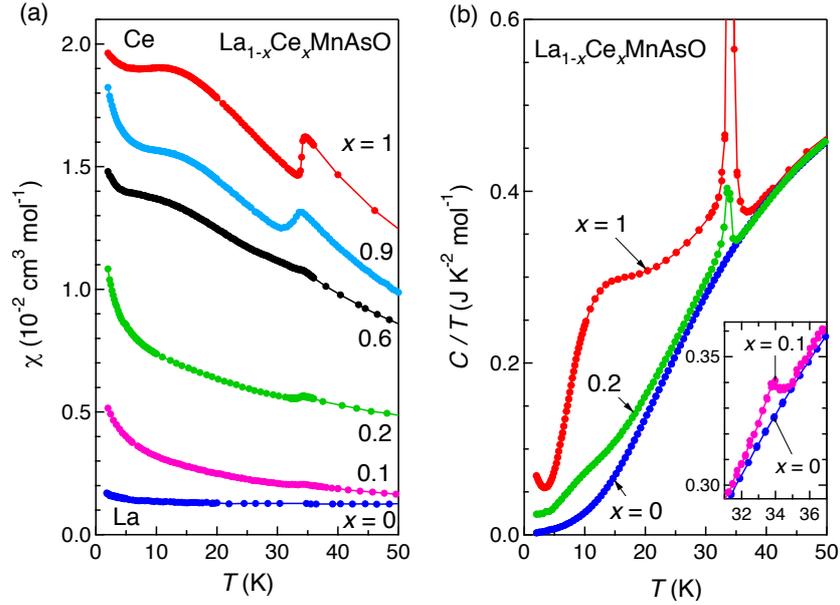

Fig. 5. (Color online) Magnetic susceptibilities (a) and heat capacities (b) of solid solutions between LaMnAsO and CeMnAsO: $La_{1-x}Ce_xMnAsO$.

In CeMnAsO, the primary origin of the transition at $T_{SR} = 34.1$ K cannot be the LRO of Ce moments through Ce-Ce interactions for three reasons. First, the temperature $T_{SR}$ is too high for such an insulator without RKKY interactions. Note that Ce moments of metallic CeFeAsO order at 4 K.[14-16] Second, if a conventional LRO sets in, magnetic susceptibility would decrease or remain constant below the ordering temperature, and heat capacity would decrease upon cooling following a power of temperature, reflecting spin-wave excitations. Instead, broad peaks are observed below $T_{SR}$ in the magnetic susceptibility and heat capacity data of CeMnAsO. Third, the phase transition survives at nearly the same temperature, even when a large number of Ce atoms are replaced by La atoms. Note that $La_{0.9}Ce_{0.1}MnAsO$ containing only 10% Ce exhibits an anomaly in the magnetic susceptibility and a small but sharp peak in the heat capacity at 34 K.

In analogy to the cases of NdMnAsO and PrMnSbO, it is plausible that the transition of CeMnAsO is also associated with the spin reorientation transition of Mn moments. As a result of the Mn moment reorientation, an internal magnetic field appears at the Ce site, splitting its ground-state doublet, just as in CeFeAsO.[16] This internal field induces Ce moments and forces them to align perfectly in a long-range manner. In this sense, the Ce moments undergo a "parasitic" order. Since no spontaneous symmetry breaking is involved in this order, the transition should be of the first order, as observed in CeMnAsO. The unusual properties of CeMnAsO, such as the increasing magnetic susceptibility upon cooling and the presence of Schottky-type heat capacity below $T_{SR}$, are ascribed to the rather small splitting of the Ce doublet, compared with that expected for a conventional LRO, which leads to a significantly large population at a higher level near $T_{SR}$ and thus to a reduced moment. The Ce moment grows more gradually upon cooling and is fully polarized at $T \rightarrow 0$ in this parasitic order. Note that, in a conventional LRO of Ce moments coupled to each other by magnetic interactions, the Ce doublet splits with a sufficiently large energy gap compared with the ordering temperature, so that a large moment is immediately induced below the transition.

It was suggested for NdMnAsO and PrMnAsO that the LROs of the Nd/Pr moments occur simultaneously at $T_{SR}$ and thus trigger the transition.[7-9] However, the temperature evolution of the Nd/Pr moments reported is quite unusual: for Nd, a steep increase at 25-20 K is followed by a gradual increase to a saturation at approximately 4 K.[7] This suggests that the moments are generated by the internal field from the reoriented Mn moments and that Nd/Pr moments also undergo parasitic ordering, as in CeMnAsO and CeFeAsO. The anomaly found in magnetic susceptibility at 4 K in NdMnAsO may show a transition to the "true" LRO of Nd moments.[7] This ordering temperature is close to those found in other related compounds such as NdFeAsO ($T_N$(Nd) = 6 K).[17] For CeFeAsO, the ground-state degeneracy is already lifted at $T_N$(Fe) = 140 K, but the energy split is so small (approximately 10 K) that the thermal population of the higher level is still large, even at low temperatures below the $T_N$. To release entropy, therefore, CeFeAsO must generate a conventional LRO at $T_N$(Ce) = 4 K through RKKY interactions. Such a need is absent in CeMnAsO, because of the large energy split of the Ce doublet (i.e., 41 K, which is comparable to $T_{SR}$), which is due probably to the larger Mn magnetic moments and stronger couplings between the Ce and Mn spins. In addition, there may be no direct interactions between the Ce moments in insulating CeMnAsO. As a result, the Ce ion falls gradually into a nonmagnetic state without showing further magnetic transitions.

For CeMnAsO, note that the high-temperature ordering of the Mn moments does not generate an internal field at the Ce site: no splitting of the Ce doublet exists above $T_{SR}$. Note that simple collinear arrangements of the Mn spins cannot produce internal fields on the Ce, because each Ce is located exactly above or below the center of the Mn square so that Heisenberg interactions cancel out at each Ce site. It has been suggested for CeFeAsO that a strong Fe-Ce coupling due to a non-Heisenberg anisotropic exchange results in large staggered Ce moments.[13] Magnetic interactions between $f$ and $d$ spins in the present family of compounds are an interesting topic for investigation in our future work,[18] and should be sufficiently strong to generate sizable internal fields at the $f$-spin site and to overcome the geometrical frustration.

The origin of the spin reorientation transition in the square lattice of Mn ions is apparently related to the presence of $f$ moments above and below the Mn lattice, because it exists in neither LaMnAsO nor BaMn$_2$As$_2$[6,9] but in NdMnAsO and PrMnSbO[7-9] and probably in CeMnAsO. We have recently studied another related compound, CeCrAsO, and observed a very similar transition at 36 K in magnetic susceptibility and heat capacity, which may be due to a spin reorientation transition. This implies that the transition is not specific to Mn ions but generally occurs in this class of compounds having both *d* and *f* moments. A spin reorientation transition has also been observed in the square lattice of Cu ions in $Nd_2CuO_4$.[19,20] There must be a certain relation between these transitions, because the local arrangements of the *d* and *f* moments are almost identical. Evidently, the issue is to clarify the interactions between *d* and *f* moments. Further investigations of these *d-f* systems would lead to the establishment of an interesting paradigm of magnetism.[18]

We have studied the magnetic properties of CeMnAsO. A first-order phase transition is found at $T_{SR}$ = 34.1 K, below which the ground-state doublet of the Ce ion suddenly splits by 3.53 meV. The transition takes place at nearly the same temperature with up to 90% substitution of La for Ce, indicating that the long-range order of Ce moments is not involved in the transition. It is suggested that Mn moments already ordered at high temperatures above 300 K are reoriented during the transition, which generates a large internal magnetic field at the Ce site through unique interactions.

**Acknowledgments**

We wish to thank S.-H. Lee and his coworkers for fruitful collaboration and helpful discussion. We are also grateful to H. Harima and K. Ueda for stimulating discussions.

Figure captions

Fig. 1. (Color online) Resistivities of polycrystalline samples of CeMnAsO, $La_{0.8}Ce_{0.2}MnAsO$, and LaMnAsO. The inset depicts the crystal structure of CeMnAsO in the ZrCuSiAs structure.

Fig. 2. (Color online) Magnetic susceptibilities $\chi$ measured at a magnetic field of 1 T upon heating polycrystalline samples of CeMnAsO and LaMnAsO. The two broken curves represent fittings to the Curie-Weiss (CW) law and were obtained by assuming crystalline electric field (CEF) splittings into three doublets for a $Ce^{3+}$ ion. The data between 100 and 300 K are used for these fittings. Inverse magnetic susceptibility is plotted in the inset.

Fig. 3. (Color online) Phase transition at 34.1 K revealed by magnetic susceptibility measured at $\mu_0 H$ = 1 T and heat capacity measured at zero magnetic field. The inset is an enlargement of the cusp observed in $\chi$ measured upon heating and cooling.

Fig. 4. (Color online) Magnetic heat capacity divided by temperature, $C_m/T$, measured at zero magnetic field for Ce ion, obtained by subtracting heat capacity of LaMnAsO from that of CeMnAsO. The solid curve represents a fit to the Schottky-type heat capacity with an energy gap of 3.53 meV. The inset schematically shows $Ce^{3+}$ CEF levels above and below the transition temperature of $T_{SR}$ = 34.1 K: the ground-state doublet splits by 3.53 meV below $T_{SR}$, and the first-excited doublet is located at 15.9 meV above $T_{SR}$, as revealed by the fitted magnetic susceptibility data shown in Fig. 2.

Fig. 5. (Color online) Magnetic susceptibilities (a) and heat capacities (b) of solid solutions between LaMnAsO and CeMnAsO: $La_{1-x}Ce_xMnAsO$.